\documentclass[prd,a4paper,superscriptaddress,nofootinbib,10pt]{revtex4} 
\usepackage{graphicx}
\usepackage{amssymb}
\usepackage{amsmath}
\usepackage{lscape}
\usepackage{mathrsfs}
\usepackage{textcomp}
\usepackage{epsfig}
\renewcommand{\d}{\mathrm{d}}

\begin{document}
\title{Noncommutative scalar quasinormal modes and quantization of entropy of a BTZ black hole}

\author{Kumar S. Gupta}
\email{kumars.gupta@saha.ac.in}
\affiliation{Theory Division, Saha Institute of Nuclear Physics, 1/AF Bidhannagar, Calcutta 700064, India}

\author{E. Harikumar }
\email{harisp@uohyd.ernet.in}
\affiliation{School of Physics, University of Hyderabad, Central University P O, Hyderabad-500046, India}

\author{Tajron Juri\'c}
\email{tjuric@irb.hr}

\author{Stjepan Meljanac}
\email{meljanac@irb.hr}
\affiliation{Rudjer Bo\v{s}kovi\'c Institute, Bijeni\v cka  c.54, HR-10002 Zagreb, Croatia}

\author{Andjelo Samsarov\footnote{On a  leave  of absence  from  the  Rudjer  Boskovic  Institute, Zagreb,  Croatia}}
\email{samsarov@unica.it}
\affiliation{Dipartimento di Matematica e Informatica, Universit\`{a} di Cagliari, viale Merello 92, 09123 Cagliari, Italy 
 and INFN, Sezione di Cagliari}

\date{\today}

\begin{abstract}

We obtain an exact analytic expression for the quasinormal modes of a noncommutative massless scalar field in the background of a massive spinless BTZ black hole up to the first order in the deformation parameter. We also show that the equations of motion governing these quasinormal modes are identical in form to the equations of motion of a commutative massive scalar field in the background of a fictitious massive spinning BTZ black hole. This results hints at a duality between the commutative and noncommutative systems in the background of a BTZ black hole. Using the obtained results for quasinormal mode frequencies, the area and entropy spectra for the BTZ black hole in the presence of noncommutativity are calculated. In particular, the separations between the  neighboring values of these spectra are determined and it is found that they are nonuniform. Therefore, it appears that the noncommutativity leads to a non-equispaced (discrete) area and entropy spectra.

\end{abstract}
%\pacs{81.05.ue, 03.65.-w}
\maketitle

\section{Introduction}

Quasinormal modes (QNM) of black holes \cite{rg,vish, cardosoreview, rew} provide an important tool to explore the AdS/CFT duality \cite{malda}. The QNM frequencies govern the decay of the gravitational perturbations in the bulk \cite{rg} and they are related to the relaxation time of the two-point function of the thermal CFT at the boundary of a BTZ black hole \cite{horo, danny1, danny2}. The QNM thus provide a correspondence between the perturbation of the gravity in the bulk to that of a boundary CFT in BTZ space-time. While this result provide an evidence for the AdS/CFT duality, it would be important to test its validity at the Planck scale, which is a natural regime for holography and quantum gravity. A first step in this direction would involve the study of QNM in a theory of gravity which is relevant at the Planck scale. Noncommutative geometry is one of the candidates of the quantum gravity at the Planck scale \cite{cones}. It is well known that the quantum theory and general relativity together lead to a noncommutative description of space-time \cite{dop1, dop2}. In particular, the noncommutative description of the BTZ black hole is given by the $\kappa$-Minkowski algebra \cite{btzkappa,ohl}. It is therefore useful to study the QNM of the BTZ black hole within the $\kappa$-Minkowski framework \cite{kappa}.

In the commutative case, the scalar QNM of the BTZ black hole are obtained by solving the Klein-Gordon (KG) equation with the QNM boundary conditions \cite{lemos,d1}. The QNM in the noncommutative case should therefore be obtained by solving the noncommutative KG equation in the BTZ background. In principle even the BTZ background could be made noncommutative. However, in this paper, we are dealing with a toy version of quantum gravity where the noncommutative effects are analyzed only up to the first order in the deformation parameter. For this purpose, it is sufficient to consider the scalar field to be noncommutative and analyze the corresponding KG equation in the background of the commutative BTZ geometry \cite{our}. 

In a previous paper we have derived the equations of motion (EOM) of a noncommutative scalar field propagating in the commutative spinless BTZ background \cite{our}. We used the $\kappa$-Minkowski algebra as a model of the noncommutativity. The noncommutative corrections to the EOM's were obtained using the method of realization \cite{sm1, sm2, sm3, epjc, twists} and are restricted to the first order in the deformation parameter. Here we will show that in terms of a suitable variable, those EOM's  reduce to an exactly solvable equation. In addition, the EOM's have a structure as those derived from a fictitious BTZ black hole with both mass and spin. In other words, the EOM of a noncommutative scalar field in the background of a massive spinless BTZ black hole have the same mathematical form as that of the EOM of a commutative scalar field in the background of a fictitious massive BTZ black hole with nonzero spin, which is exactly solvable. Both the mass and spin of this fictitious black hole contain terms which are proportional to the deformation parameter. This allows us to obtain an analytical expression for the QNM of the noncommutative scalar field in the BTZ background. The QNM thus obtained contain the noncommutative corrections up to the first order and have the correct commutative limit. This toy model therefore gives a hint as to how the Planck scale physics might affect the QNM frequencies.

There is an interesting proposal that relates the QNM frequency to the black hole area quantization. The area quantization was originally proposed by Bekenstein \cite{bekenstein} assuming that the area of the black hole horizon is an adiabatic invariant. Hod proposed that the level spacing of the quantized area can be determined from the real part $\mbox{Re} \; \omega $ of the highly damped QNM frequencies \cite{Hod:1998vk}. Subsequently it was proposed by Maggiore \cite{Maggiore:2007nq} that the level spacing of the quantized black hole area should be related to
% $\sqrt{{Re\omega}^2 + \omega_I^2}$
 $\sqrt{{(\mbox{Re} \; \omega)}^2 + {(\mbox{Im} \; \omega)}^2 },$
  where $\mbox{Im} \; \omega$ is the imaginary part of the QNM frequency. Here we shall use the proposal by Maggiore to obtain a noncommutative correction to the black hole area law quantization.

 The plan of the paper is the following.
  In section II we investigate the radial equation of motion governing the noncommutative (NC) scalar field in the background of the massive spinless BTZ black hole. We find  quasinormal mode solutions to this equation, that is wave functions and corresponding energies/frequencies. In  section III we show that the physical system described by the radial equation from the preceding section is mathematically equivalent to the physical system comprised of a massive commutative scalar field in a background of the massive spinning BTZ black hole. Following this, we give a novel view on noncommutativity, which emerges from our analysis, by assigning it a role of a mass generating agent, as well as a driving force that lies behind the black hole spin (see section IV). In section IV the issue of the area and entropy quantization is discussed. It is shown that the spectra of both of these quantities are discrete and nonequidistant. Moreover, it appears that the source responsible for the none
  equidistant nature of the spectra is the spacetime noncommutativity, which apparently breaks the equidistantness in the area/entropy spectrum, so that their eigenvalues are no  more equally spaced. We conclude with section 5.

%%%%%%%%%%%%%%%%%%%%%%%%%%%%%%%%%%%%%%%%%%%%%%%%%%%%%%%%%%%%%%%%%%%%%%%%%
%%%%%%%%%%%%%%%%%%%% %%%%%%%%%%%%%%%%%%%%%
%%%%%%%%%%%%%%%%%%%%%%%%%%%%%%%%%%%%%%%%%%%%%%%%%%%%%%%%%%%%%%%

\section{Quasinormal modes}

In \cite{our} we have investigated some of the properties of $\kappa$-deformed scalar field theory in the curved background. As explained before, we consider a commutative geometry which is probed with a NC scalar field. The noncommutativity is introduced by replacing the usual pointwise multiplication, between the fields in the action functional, with the NC star product, i.e. $\phi(x)\phi(x)\longrightarrow\phi(x)\star\phi(x)$. Then,  we postulated the following physical action 
\begin{equation}\begin{split}\label{S1}
\hat{\mathcal{S}}&=\int \text{d}^{4}x\sqrt{-g}\ \ g^{\mu\nu}\left(\partial_{\mu}\phi\star\partial_{\nu}\phi\right)\\
&=\int \text{d}^{4}x\sqrt{-g}\ \ g^{\mu\nu}\left(\partial_{\mu}\hat{\phi}\partial_{\nu}\hat{\phi}\triangleright 1\right)\\
&=\mathcal{S}_{0}+\int\text{d}^{4}x\left(\mathcal{A}_{\rho\sigma\gamma\delta}\frac{\partial^{2}\phi}{\partial x_{\rho} \partial x_{\sigma}}\frac{\partial^{2}\phi}{\partial x_{\gamma} \partial x_{\delta}}\right)+O(a^2)
\end{split}\end{equation}
where we used  the ``method of realization'' \cite{sm1, sm2, sm3, epjc, twists}, $\mathcal{S}_{0}$ is the undeformed action and $\mathcal{A}_{\rho\sigma\gamma\delta}$ is given by\footnote{In eq.\eqref{A} $a_\mu$ is a 4-vector of deformation, but in the subsequent analysis we choose one particular orientation, $a_\mu = (a,0,0,0)$, so that henceforth the symbol $a$ is reserved for the time component of the deformation 4-vector.  }
\begin{equation}\label{A}
\mathcal{A}_{\rho\sigma\gamma\delta}=i\sqrt{-g}\ g_{\sigma\delta}\left(\alpha x_{\rho}a_{\gamma}+\beta(a\cdot x)\eta_{\rho\gamma}+\gamma a_{\rho}x_{\gamma}\right).
\end{equation}
% % % % % % % % % % % % %
Here $\alpha, \beta, \gamma$ are the parameters determining the differential operator representation of the $\kappa$-Minkowski algebra \cite{kappa}. In what follows we focus on the family of realizations that is parametrized only by the parameter $\beta$ (since only terms proportional to $\beta$ appear in the equations of motion in the long wavelength approximation \cite{our}). In this family of realizations there is one that is particularly interesting. It is determined by $\beta =1$ and corresponds to the phase space noncommutativity that is related to a generalized uncertainty principle emerging from a study of string collisions at Planckian energies \cite{gross}. It is for the first time considered by Maggiore \cite{genunc}. It was also considered in \cite{sm2, epjc, kowalski, jonke, borowiecpachol, sams} where it is denoted as natural realization or classical basis.
% % % % % % % % % % % %

 Eq. \eqref{S1} is valid for a general curved spacetime metric $g_{\mu\nu}(x)$. For our analysis, we shall focus on the commutative massive BTZ geometry \cite{banados}, which is taken to be spinless. We substitute this BTZ metric explicitly in Eq. \eqref{S1} and use the $\kappa$-deformed
scalar field to probe the geometry. The massive spinless BTZ black hole is described by the metric \cite{banados}
\begin{equation}\label{btzmetric}
g_{\mu\nu}=\begin{pmatrix}
\frac{r^2}{l^2}-M&0&0\\
0&-\frac{1}{\frac{r^2}{l^2}-M}&0\\
0&0&-r^2\\
\end{pmatrix},
\end{equation}
where we have taken  the angular momentum to be zero, i.e.  
$J=0$ and  $l$ is related to the cosmological constant $\Lambda$ as $l = \sqrt{-\frac{1}{\Lambda}}$. In order to find the $\kappa$-deformed Klein-Gordon equation, for simplicity, we use the long-wavelength limit and obtain the following 
radial equation \cite{our}
\begin{equation}\label{eomradial}
r\left(M-\frac{r^2}{l^2}\right)\frac{\partial^2 R}{\partial r^2}+\left(M-\frac{3r^2}{l^2}\right)
\frac{\partial R}{\partial r}+\left(\frac{m^2}{r}-\omega^2\frac{r}{\frac{r^2}{l^2}-M}-a\beta\omega\frac{8r}{l^2}\frac{\frac{3r^2}{2l^2}-M}{\frac{r^2}{l^2}-M}\right)R=0,
\end{equation}
which will be the cornerstone of the whole subsequent analyzes, presented in this paper. See Appendix A for a brief derivation of \eqref{eomradial} whose full details can be found in \cite{our}.

Before proceeding further, let us address an important point. The NC field theory is highly non-linear and it may therefore appear surprising that the higher derivative terms do not contribute to the equation of motion. This point has been discussed in detail in \cite{our} and here we shall recall the essential features of that analysis. The main point to note is that we are looking for the NC correction to the lowest order in the deformation parameter. The NC effects are expected to arise at the Planck scale and the NC deformation parameter would be suppressed in powers of the Planck mass. It is therefore logical to consider the NC effects only to the lowest order. In addition, we also look at the long wavelength or low frequency limit of the quasi-normal modes. The reason is that the QNM's are associated to the gravitational perturbations, which are inherently very weak \cite{kokkotas} and there is a considerable effort from the experimental side to detect the low frequency signals \cite{cam}. It is therefore both logical and important to consider the low frequency limit. With these two approximations, the equation of motion reduces to the form given in \eqref{eomradial}. 

Now we  show that eq. \eqref{eomradial} has an exact solution for QNM boundary conditions.
Using the following substitution
\begin{equation}
z=1-\frac{Ml^2}{r^2},
\end{equation}
we re-express eq.\eqref{eomradial}  as
\begin{equation}
\label{eom}
z(1-z)\frac{\d^2 R}{\d z^2}+ (1-z)\frac{\d R}{\d z} + \left(\frac{A}{z}+B+\frac{C}{1-z} \right)R=0,
\end{equation}
where the constants $A,B$ and $C$ are
\begin{equation} \label{coefs}
A=\frac{\omega^2 l^2}{4M}+a\beta\omega, \quad B=-\frac{m^2}{4M}, \quad C=3a\beta\omega.
\end{equation}

Radial eq.\eqref{eom} is analytically solvable\footnote{Note that eq.\eqref{eom} has the same form as eq.(8) in \cite{d1}.} and the solution is given by
\begin{equation}
R(z)=z^{\lambda_1}(1-z)^{\lambda_2}F(z)
\end{equation}
where $F(z)$ is the hypergeometric function satisfying
\begin{equation}
z(1-z)\frac{\d^2 F}{\d z^2}+\left[c-(1+a+b)z\right]\frac{\d F}{\d z}-abF=0.
\end{equation}
After defining
\begin{equation}
c=2\lambda_1+1, \quad a+b=2\lambda_1+2\lambda_2,  \quad ab=(\lambda_1+\lambda_2)^2 -B
\end{equation}
and
\begin{equation}
\lambda_1^{2}=-A, \quad \lambda_2=\frac{1}{2}(1\pm\sqrt{1-4C}),
\end{equation}
 it subsequently follows
\begin{equation}
a=\lambda_1+\lambda_2+i\sqrt{-B}, \quad b=\lambda_1+\lambda_2-i\sqrt{-B}.
\end{equation}
We choose $\lambda_1=-i\sqrt{A}$ and $\lambda_2=\frac{1}{2}(1-\sqrt{1-4C})$ without loss of generality. 

The quasinormal modes are defined as solutions which are purely ingoing at the horizon, and which vanish at infinity \cite{horo}. We have two linearly independent solutions of eq.\eqref{eom},
$F(a,b,c,z)$ and $z^{1-c}F(a-c+1,b-c+1,2-c,z)$ near the horizon $z=0$. Thus, the solution which has ingoing flux at the horizon is given by
\begin{equation}\label{R}
R(z)=z^{\lambda_1}(1-z)^{\lambda_2}F(a,b,c,z)
\end{equation}
Since \eqref{R} is valid only in some neighborhood of the horizon, for the infinity, $z=1$, we use the linear transformation formula
\begin{equation}\begin{split}
R(z)=&z^{\lambda_1}(1-z)^{\lambda_2+c-a-b}\frac{\Gamma(c)\Gamma(a+b-c)}{\Gamma(a)\Gamma(b)}F(c-a,c-b,c-a-b+1,1-z)\\
&+z^{\lambda_1}(1-z)^{\lambda_2}\frac{\Gamma(c)\Gamma(c-a-b)}{\Gamma(c-a)\Gamma(c-b)}F(a,b,a+b-c+1,1-z),
\end{split}\end{equation}
where the first term vanishes and the vanishing of the second term imposes the following conditions
\begin{equation}
c-a=-n, \quad \text{or} \quad c-b=-n,
\end{equation}
and $n=0,1,2...$  These conditions determine the frequencies of the quasinormal modes. The left and right quasinormal mode 
frequencies
are given by\footnote{We neglect the terms of the order $O(a^{2}).$}
\begin{equation}\label{freq}
\omega_{L,R}=\pm\frac{m}{l}+a\beta\frac{2M}{l^2}(6n+5)-2i\left[\frac{\sqrt{M}}{l}(n+1)\mp3a\beta\frac{m}{l^2}\sqrt{M}\right].
\end{equation}
 Notice that for $a\rightarrow 0$ we recover the result for the quasinormal frequencies of massless scalar field in the BTZ background with $J=0$ (see \cite{d1}). Also it may be observed that noncommutativity introduced the $n$ dependence into the real part and the angular momentum $m$ dependence into the imaginary part of the quasinormal mode frequencies.

\section{Mapping to a fictitious commutative BTZ black hole}

The equation \eqref{eom}, together with the coefficients \eqref{coefs}, describes the dynamics of massless NC scalar field with energy $\omega$ and angular momentum $m$, in the background of a BTZ black hole with mass $M$ and with vanishing angular momentum ($J=0$). A close inspection of the form of 
equation \eqref{eom} leads to an interesting observation. To see that, consider the equation of motion of a different scalar field of mass $\mu^f$, energy $\omega$ and angular momentum $m$ in the background of a fictitious BTZ black hole with mass $M^f$ and angular moment $J^f$, which corresponds to equation (8) of ref.\cite{d1}. Our equation \eqref{eom} and equation (8) of ref.\cite{d1} have exactly the same analytical structure, except that the coefficients $A, B$ and $C$ in ref.\cite{d1} have different forms, which depend on the parameters of the new scalar field and the fictitious BTZ black hole. Thus we see that the equations of motion of a massless noncommutative scalar field in the background of a massive spinless BTZ black hole has the identical form as that of a massive commutative scalar field in the background of a massive spinning BTZ black hole. 

Since we have the equivalence between two equations of motion, which pertain to the two completely different physical situations, a question naturally arises as to whether it is possible to find some kind of mathematical correspondence between them.The answer is yes. Namely,
it appears that it is really possible  to
    find a direct mapping between the case considered here, that is NC massless scalar field in the non-rotational BTZ background, and
     the physical setting where the ordinary massive scalar field probes a BTZ geometry with nonvanishing angular momentum. In what follows, the later setting we shall refer to  as the fictitious one (see \cite{d1} for the explicit expressions that correspond to this fictitious situation).
                
                Therefore, 
      by comparing the constants $A,B$ and $C$, appearing in \eqref{eom}, with the appropriate constants from the reference \cite{d1}, we get the following set of conditions
\begin{equation}\begin{split}
&A=\frac{\omega^2 l^2}{4M}+a\beta\omega=\frac{l^4}{4(r^2_{+}-r^2_{-})^2}\left(\omega r_{+}-\frac{m}{l}r_{-}\right)^2=A^f\\
&B=-\frac{m^2}{4M}=-\frac{l^4}{4(r^2_{+}-r^2_{-})^2}\left(\omega r_{-}-\frac{m}{l}r_{+}\right)^2=B^f\\
&C=3a\beta\omega=-\frac{\mu^f}{4}=C^f,
\end{split}\end{equation}
with $\; r_+, r_- \;$ being the outer, i.e. inner radius of the equivalent spinning BTZ black hole, respectively\footnote{Note that to be completely correct, we should use the notation $r^f_+$ and $r^f_-$ instead of $r_+, r_-$ for the radii, to distinguish them from the radii pertaining to the spinless BTZ, which are
respectively given by $r_+ = l\sqrt{M}$ and  $r_- =0 $. However, to keep the notation as simple as possible, we omitted the superscript.}. 

Furthermore, since \cite{banados}
\begin{equation}
M^f=\frac{r^{2}_{+}+r^{2}_{-}}{l^2}, \quad  J^f=\frac{2r_{+}r_{-}}{l},
\end{equation}
we can express the parameters of the commutative fictitious situation completely in terms of the parameters defining the NC case  we analyze here,
\begin{equation}
M^f=M^f(a,M), \quad J^f=J^f(a,M), \quad \mu^f=\mu^f(a,M).
\end{equation}
This mapping is similar to the one obtained in \cite{veza}, where the analogy between the NC version of Schwarzschild black hole and the commutative Reisner-Nordstrom black hole was made.

In order to understand the physical meaning behind the above equivalence, note that when the noncommutative parameter goes to zero, the parameters of the two situations coincide with each other. This observation suggests that the noncommutativity of the scalar field generates, possibly through some back reaction, the additional mass and angular momentum of the system with the fictitious black hole.  
 
\section{Quantization of entropy}

In  quantum gravity theory the black hole area is represented by a quantum operator and  its values are therefore 
supposed to constitute a discrete set, made of eigenvalues of the area operator. In order to gain  insight into the 
nature of the area spectrum, one may approach the problem in a semiclassical way, which utilities the Bohr-Sommerfeld 
quantization applied to a typical adiabatic invariant, characteristic for the physical system in question. Generally, 
for the system of energy $E$ and characteristic frequency $\Delta \omega (E)$, the typical adiabatic invariant is of 
the form $\dfrac{E}{\Delta \omega (E)}$ \cite{Kunstatter:2002pj}. In our case of interest, namely the black hole with  horizon 
area, we may replace the energy $E$ with the black hole mass $M$ and for  $\Delta \omega (E)$ we may take 
\cite{vagenas},\cite{medved} the transition frequency between the two adjacent states of the black hole. The qualitative picture
one may have in mind here is the semiclassical one where the black hole is considered as a physical system that can 
exist in different quantum states and whose energies are given by the quasinormal mode frequencies.

When undergoing a transition between the two states in their quasinormal mode spectrum,
 these oscillating black holes then emit elementary quanta of energy/mass.
    However, since quasinormal mode frequencies are complex valued quantities, a question naturally raises
     as to which part of them  carries the genuine information about the energy, especially the elementary 
     quantum of energy/mass that is emitted or absorbed from outside of the black hole horizon.
     
       In this respect, the first who pointed out that the quasinormal modes could have a relevance to these 
       elementary quanta and subsequently to the quantization of the area and entropy of the black hole was Hod \cite{Hod:1998vk}.
       He identified the mass of the elementary black hole quanta with the real part of the quasinormal mode 
       frequencies. Afterwords, while retaining the crucial role for the quasinormal modes in quantizing the black 
       hole area, Maggiore  proposed that instead of the real part of the quasinormal mode frequencies, it might be 
       that it is their absolute value that has the real physical meaning \cite{Maggiore:2007nq}.
       
        If we apply this conjecture to our case
        in hand, we get that the elementary quantum of mass $\Delta M$ that is emitted or absorbed by the black hole is given by 
        \begin{equation}\label{Mag}
        \Delta M = \hbar \Delta \omega = \hbar (\left|\omega_{L,R}\right|_{n}-\left|\omega_{L,R}\right|_{n-1}),
       \end{equation}
       where $\left|\omega_{L,R}\right| = \sqrt{{(\mbox{Re} \; \omega_{L,R})}^2 + {(\mbox{Im} \; \omega_{L,R})}^2 }$ and $n$ is generally understood to be large, in concordance with the Bohr's correspondence principle which only holds for transitions where both radial quantum numbers (here $n$ and $n-1$) are large. Nevertheless, even if we extrapolate the analysis to low $n$ and to the transitions from and to a black hole in its fundamental state, where semiclassical reasoning might be wrong, we are still left with a non-vanishing quantum of mass  (i.e. quantum of black hole area). 
       
Now, using quasinormal mode frequencies \eqref{freq}, the frequency difference 
$\Delta\omega$ can be calculated as
 \begin{equation}\label{delta}
    \Delta\omega=\left|\omega_{L,R}\right|_{n}-\left|\omega_{L,R}\right|_{n-1}=\frac{2\sqrt{M}}{l}\left(1\pm \frac{a \beta}{2l}\frac{m}{n(n+1)}\right).
\end{equation}
Recalling that the area of the non-rotating BTZ black hole is given by 
\begin{equation} \label{NCBTZarea}
A=2\pi r_{+}=2\pi l\sqrt{M},
\end{equation}
we determine the elementary quantum $\Delta A = 2 \pi l\frac{1}{2\sqrt{M}}\Delta M$ of the BTZ black hole area, in the presence of noncommutativity, to be given as
\begin{equation} \label{deltaA}
 \Delta A= 2\pi \hbar \left(1\pm \frac{a \beta}{2l}\frac{m}{n(n+1)} \right).
\end{equation}
The result obtained suggests that the amount of area by which the black hole horizon "shrinks"
depends on the quantum states of the black hole that are involved in the particular process of transition.
%  which quantum states of the black hole are %involved in the process of transition.
One may note that in the limit of vanishing noncommutativity, $a \rightarrow 0$, this feature disappears, keeping the elementary quantum of area insensitive to the details of the quasinormal mode spectra (or equivalently, the black hole spectra).

  In order to determine, within the semiclassical approach, the area spectrum for the BTZ black hole in noncommutative setting, we turn to the observation mentioned at the beginning of this section in relation to adiabatic invariant. As already noted, for the black hole of mass $M$ and characteristic vibrational  frequency $\Delta \omega$ the quantity of interest is
\begin{equation}\label{I}
\mathcal{I}=\int \frac{\d M}{\Delta \omega},
\end{equation}
 where for the large $n$ the vibration frequency $\Delta\omega$ is given by \eqref{delta}. In making this choice here we follow \cite{vagenas},\cite{medved},\cite{Wei:2009yj}.

 Then again, by using quasinormal mode frequencies \eqref{freq} and substituting \eqref{delta} in \eqref{I},
 it is possible to evaluate the adiabatic invariant $\mathcal{I}$ as
\begin{equation}
\mathcal{I}=l\sqrt{M}\frac{1}{1\pm \frac{a \beta}{2l}\frac{m}{n(n+1)}}\approx l\sqrt{M}\left(1\mp \frac{a \beta}{2l}\frac{m}{n(n+1)} \right).
\end{equation}
Now, imposing the Bohr-Sommerfeld quantization condition, we have
\begin{equation}
\mathcal{I}=N \epsilon, \quad N\in\mathbb{N},
\end{equation}
where, unlike for example \cite{Wei:2009yj}, we allow a somewhat broader range of possibilities for $\epsilon$.  It might be $\hbar$, Planck length or some intermediate scale $a$, which settles the noncommutativity scale somewhere in between LHC energies and the Planck energy.
 %Shall we just put $\epsilon=\hbar$, 
Based on the formula \eqref{NCBTZarea},
 the spectrum of the area of BTZ black hole in the setting of noncommutative geometry 
 is
\begin{equation} \label{finalarea}
A=2\pi\mathcal{I}\left(1\pm \frac{a \beta}{2l}\frac{m}{n(n+1)}\right)\Rightarrow A_{N}=2\pi N\epsilon \left(1\pm \frac{a \beta}{2l}\frac{m}{n(n+1)} \right).
\end{equation}
Note that this result is in concordance with \eqref{deltaA} since it predicts the same elementary quantum of area, although obtained by somewhat different approach \cite{Maggiore:2007nq}. It can also be seen that
while the spectrum is linear in the quantum number $N$, it becomes increasingly more complex in the quantum numbers $n,m$. Furthermore, one finds that it is the effect of noncommutativity that brings in this additional richness and complexity into the pattern of the area spectrum, rendering it nonuniform. This feature is new when compared to \cite{Wei:2009yj}.
Note however that for $a \rightarrow 0,$ our result reduces to that in \cite{Wei:2009yj}.

Assuming the Bekenstein-Hawking relation, the entropy of the nonrotational BTZ black hole in the same noncommutative setting is then given by 
\begin{equation} \label{finalentropy}
S_{N}=\frac{A_{N}}{4G^*},
\end{equation}
where {\footnote{Note that so far we were carrying the study in the units where $8G = 1$. For the purpose of the remaining analysis we switch to the standard unit system, which in turn formally corresponds to putting $8GM$ everywhere in place of $M$. Note also that the result \eqref{finalarea} is insensitive to this change. That is, it remains the same after one makes the above described change of units.}} $G^*$ denotes
the rescaled Newton's gravitational constant, with rescaling being induced by the noncommutativity related effects (see \cite{our}
for more details). Utilizing in  \eqref{finalentropy} the result for the rescaled gravitational constant from \cite{our}, we get for the entropy
\begin{equation} \label{finalentropy1}
S_{N}=\frac{A_{N}}{4G} \bigg(1+ a \beta \dfrac{8}{3} \dfrac{\pi}{l} \dfrac{\zeta(2)}{\zeta(3)} \sqrt{8GM}  \bigg),
\end{equation}
which  after making use of the area relation
\eqref{finalarea} and keeping only terms linear in $a$, subsequently leads to the result
\begin{equation} \label{finalentropy2}
S_{N}= N\epsilon \frac{\pi}{2G} \bigg(1 \pm \frac{a \beta}{2l} \frac{m}{n(n+1)}+ a \beta \dfrac{8}{3} \dfrac{\pi}{l} \dfrac{\zeta(2)}{\zeta(3)} \sqrt{8GM}  \bigg).
\end{equation}
 We see from this result that the entropy exhibits the similar pattern of behavior as its area counterpart. 
  % % % % % % % % % % % % % %
 Thereby the noncommutativity breaks the equidistantness in the spectrum. The horizon area, being quantized with elementary quantum
 of  area \eqref{deltaA}, must be of the form $A_N= N\Delta A $. Hence, at the level $N$ (and with the choice $\epsilon=\hbar$) one can expect that the number of black hole microstates is given by $B_N = \exp(N/(4G) 2\pi \hbar (1 \pm \frac{a \beta}{2l} \frac{m}{n(n+1)}))$.
 % % % % % % % % % % % % % % %
 
 Besides, the interesting point about the entropy \eqref{finalentropy2}
 is that it may be used to deduce the angular momentum of the equivalent spinning black hole.
 We may recall the argumentation from section III where we argued that the nonrotational BTZ black hole in noncommutative setting can be mapped to an equivalent spinning BTZ in the absence of noncommutativity.
 The point here is that the above result may be used to deduce the angular momentum which pertains to this equivalent picture and may be viewed as being generated due to the effects driven by noncommutativity. Bearing on the equivalence between two pictures, one may equate the corresponding entropies, namely $ S=S^f $. Here $S$ is the entropy \eqref{finalentropy2} and $S^f = \frac{A^f}{4G}$ is the entropy of the equivalent spinning BTZ black hole in the standard commutative background. Owing to the fact that the later case describes the spinning black hole, the corresponding radius $r_+^f$  acquires the additional contribution due to the nonvanishing angular momentum
 \begin{equation} \label{comradius}
 r_{+}^f=\frac{l\sqrt{8GM}}{2} \bigg(1+ \sqrt{1- \bigg( \dfrac{J^f}{8GMl}\bigg)^2}  \bigg),
 \end{equation}
 so that the corresponding area $A^f = 2\pi r_+^f$ modifies appropriately. Of course, since in the later case the underlying setting is  commutative one, one may use the standard gravitational constant $G$ to calculate the entropy.
  The straightforward calculation then shows that the square of the induced angular momentum scales with the noncommutativity parameter as
\begin{equation} \label{inducedangmom}
 {J^f}^2 \sim  a \beta N^2 {\epsilon}^2
    \dfrac{\pi}{l} \dfrac{\zeta(2)}{\zeta(3)} {(8GM)}^{3/2} + {\mathcal{O}} (a^2),
 \end{equation} 
  exposing in this way the black hole angular momentum as
 being purely due to the noncommutative nature of the underlying spacetime geometry.  

 Finally, few comments concerning the decay rate of a black hole may be given. 
  The decay of small perturbations of a black hole at equilibrium may be described by the quasinormal modes, which represent a discrete set of solutions to the
 wave equation subject to particular boundary conditions, and which are characterized by
   the spectrum of complex frequencies.
 
  In the most rudimentary approximation of the black hole, which may be taken to be
 that of the classical damped oscillator,  the imaginary part of   the quasinormal mode frequencies may be related to the decay rate of the black hole, or equivalently, to the relaxation of the
  system back to thermal equilibrium \cite{danny1},\cite{Maggiore:2007nq}.
  In this picture the small perturbations of black holes  vanish in time as a
superposition of damped oscillations
%with a spectrum of complex frequencies
whose decay time is then determined by the imaginary part 
 of these quasinormal frequencies. 
 
  From the result \eqref{freq} obtained in section II
   we see that the imaginary part of the  QNM frequencies scales linearly with the radial
    quantum number $n$, which is seen to be the case for both, left and right modes. This means that as $n$ grows, which corresponds to more and more excited states, the imaginary part of the QNM frequencies also grows. Moreover, as their
   imaginary part is proportional to the decay rate, which in turn is inversely
  proportional to the life time of the particular excited state, we finally
  have the conclusion that higher excited states live shorter. Since this result is highly plausible on the general ground  of physical intuition, we stress that our calculations confirm and support this assertion.
 
 The next interesting observation is that   
 the effect of noncommutativity is either to suppress or boost the decay rate depending on whether the left or right modes are concerned.
  In particular, as $n$ increases, the decay rate will increase too, but for the left modes this effect will be suppressed,  due to the presence of noncommutativity.
  Analogously, for the right modes  the increase of the decay rate (decrease in the life time) with the increase of $n$ will be additionally boosted, that being the net effect of noncommutativity too, as is plainly suggested from the expression \eqref{freq}.

\section{Conclusion}

In this paper we have calculated the noncommutative corrections to the QNM frequencies of the BTZ black hole, up to the first order in the deformation parameter.  The calculation has been performed within the context of the $\kappa$-Minkowski algebra, which is a type of noncommutativity that is associated with several black hole spacetimes. We started with a massless NC scalar field coupled to a commutative spinless BTZ background, from which
we obtained the differential equation and deformed QNM were obtained by solving this equation.
%which is solved to obtained the deformed QNM. 
This equation has the same analytic form  as that of a massive commutative scalar field in the background of a massive spinning commutative BTZ black hole. This remarkable result, which is valid up to the first order in the deformation parameter, allows us to analytically obtain the deformed NC QNM. The deformed NC QNM frequencies reduce to their commutative counterpart when the deformation parameter is removed.

This result indicates that the NC deformation of the field operator in the lowest order respects the $SL(2,R)$ symmetry of the BTZ black hole, which can be understood as follows. Let us first note that we started with a classical commutative BTZ black hole and considered a NC scalar field in its background. At that level, the classical symmetries of the BTZ are left unchanged. But still there is a question if the NC effect somehow spoils it. It has been shown in an earlier paper \cite{btzkappa} that general deformations of the BTZ black hole are of the $\kappa$-Minkowski type. Indeed, the way this was deduced in \cite{btzkappa} is by asking what is the most general type of NC deformation that is compatible with the symmetries of the BTZ and that led to a $\kappa$-Minkowski type algebra. This has also been subsequently found in other works, namely that in \cite{ohl}. Hence, at the NC level, $\kappa$-Minkowski algebra is compatible with the symmetries of the BTZ black hole. In the present work, we have a dual scenario where the field is NC and the geometry is classical. In this dual picture, and in the lowest order in the deformation parameter as mentioned above, the NC field respects the symmetries of the BTZ space-time. The terms which are of higher orders in the NC parameter, including the ones containing the higher derivatives, are in general unlikely to respect the classical symmetries of the BTZ black hole.  

As discussed in \cite{horo,danny1,danny2}, QNM of the BTZ black hole provide a tool to study AdS/CFT duality or holography. In addition, the commutative BTZ black hole is compatible with a precise kinematical description of holography, as encoded in the Sullivan's theorem \cite{sullivan}. This theorem states that for 
certain classes of hyperbolic spaces, which are called geometrically finite, there is a one-to-one correspondence 
between the hyperbolic structure in the interior and the conformal structure at the boundary. It has been shown that 
the Euclidean BTZ black hole as well as some of its variants are geometrically finite for which the Sullivan's theorem holds \cite{sendanny,senksg1,senksg2}. We have already mentioned 
that our resulting equation for the QNM has the structure as if it was obtained for a massive commutative scalar 
field in the background of a commutative massive spinning BTZ black hole. Note that Sullivan's theorem is valid for 
these classes of black holes. We can therefore say that the AdS/CFT duality or the holographic principle is compatible 
with the probing of a BTZ black hole geometry with a noncommutative scalar field obeying the $\kappa$-Minkowski algebra. 

It is natural to ask how general our results would be, since they are specific to the BTZ blak hole. In this context it may be noted that a very large class of string theories contain a BTZ factor in the near-horizon limit \cite{malda}. It is thus plausible that our results would apply for this wide class of geometries as well (even for dilaton black holes \cite{dil}). 

 Following the idea due to Bekenstein that in quantum gravity the area of the black hole horizon should be 
 quantized, we turned to the problem of finding the discrete spectrum for the BTZ black hole horizon area and the corresponding entropy. 
  In the approach presented here we followed the original proposal put forward by Hod that
  QNM, as the leading and most dominant signal from the gravitational perturbations of the oscillating black holes, may shed some light on the problem of the black hole area quantization. This in particular pertains to the spacing of the area spectrum, for which varied results are available in the literature. Here 
  we find that within the $\kappa$-deformed noncommutative framework, the area and entropy of the BTZ black hole  are both described by the discrete spectra comprising of the sets of nonequidistant values.
  
   It should be noted that this conclusion differs from the results known  in the literature \cite{Wei:2009yj}, where the BTZ black hole spectra for the area as well as for the entropy were found to be equidistant.
    Here we have a different situation, namely, based on the results  obtained here, it appears that it is the noncommutativity which makes the difference and gives rise to a nonuniform separation between the  adjacent values across the area and entropy spectra.
    Moreover, the bare effect of the noncommutativity is to give the relevance to both quantum numbers, the radial quantum number and the angular momentum quantum number, which are otherwise put aside,
     and intertwine them to produce the nonequidistant spectra. In the absence of noncommutativity these
     quantum numbers do not play  any role in shaping the area/entropy spectrum and are completely irrelevant when it comes up to an identification of the  elementary quanta for these quantities (black hole area/entropy). As we have seen, this situation  abruptly changes once the noncommutativity is switched on.\\

%%%%%%%%%%%%%%%%%%%%%%%%%%%%%%%%%%%%%%%%%%%%%%%%%%%%%%%%%%%%%%
%%%%%%%%%%%%%%%%%%%%%%%%%%%%% 
%%%%%%%%%%%%%%%%%%%%%%%%%%%%%%%%%%%%%%%%%%%%%%%%%%%%%%%%%%%%%%%

\noindent{\bf Acknowledgment}\\
 A.S. is grateful to S.Mignemi for extensive discussions and many useful comments. The work of T.~J. and S.~M.  has been fully supported by Croatian Science Foundation under the project (IP-2014-09-9582). The work of A.~S. was supported by the European Commission and the Croatian Ministry of Science, Education and Sports through grant project financed under the Marie Curie FP7-PEOPLE-2011-COFUND, project NEWFELPRO.

\appendix
\section{$\kappa$-deformed Klein-Gordon equation in the BTZ background}

 We start with the following physical action for the NC scalar field 
\begin{equation}\label{S11}
\hat{\mathcal{S}}=\int \text{d}^{4}x\sqrt{-g}\ \ g^{\mu\nu}\left(\partial_{\mu}\phi\star\partial_{\nu}\phi\right).
\end{equation}
Now we shall use  the ``method of realization'' \cite{sm1, sm2, sm3, epjc, twists} up to the first order in $a$, where the star product is given by
\begin{equation}\label{star1}
f(x)\star g(x)=f(x)g(x)+i\alpha(x\cdot \frac{\partial f}{\partial x})(a\cdot \frac{\partial g}{\partial x})+i\beta(a\cdot x)(\frac{\partial f}{\partial x}\cdot \frac{\partial g}{\partial x})+i\gamma(a\cdot \frac{\partial f}{\partial x})(x\cdot \frac{\partial g}{\partial x}).
\end{equation}
Setting $f=g=\partial\phi$ in Eq. \eqref{S11}, we expand the action up to the first order 
in the deformation parameter $a_{\mu}$ as
\begin{equation}\label{akcija}
\hat{\mathcal{S}}=\mathcal{S}_{0}+\int\text{d}^{4}x\left(\mathcal{A}_{\alpha\beta\gamma\delta}\frac{\partial^{2}\phi}{\partial x_{\alpha} \partial x_{\beta}}\frac{\partial^{2}\phi}{\partial x_{\gamma} \partial x_{\delta}}\right),
\end{equation}
where we define
\begin{equation}\label{aa}
\mathcal{A}_{\alpha\beta\gamma\delta}=i\sqrt{-g}\ g_{\beta\delta}\left(\alpha x_{\alpha}a_{\gamma}+\beta(a\cdot x)\eta_{\alpha\gamma}+\gamma a_{\alpha}x_{\gamma}\right).
\end{equation}

Starting from the noncommutative scalar field theory described by the
above action, we derive the equations of motion for the 
field $\phi$. Notice that the action in Eq.(\ref{akcija}) has the terms involving higher derivatives of the scalar 
field, i.e., our Lagrangian is $\mathcal{L}=\mathcal{L}(\phi,\partial \phi, \partial^{2}\phi, x)$  and hence the
Euler-Lagrange equations will be more general,  as in the case of higher derivative theories. Therefore, the 
Euler-Lagrange equation relevant here is
\begin{equation}\label{EL}
\partial_{\mu}\frac{\delta \mathcal{L}}{\delta(\partial_{\mu}\phi)}-\partial_{\mu}\partial_{\nu}\frac{\delta \mathcal{L}}{\delta(\partial_{\mu}\partial_{\nu}\phi)}=\frac{\delta \mathcal{L}}{\delta \phi}.
\end{equation}
Before calculating the Euler-Lagrange equation let us notice that $\frac{\partial^2 \phi}{\partial x_{\alpha}\partial x_{\beta} }=(\partial^{\alpha}g^{\beta\rho})\partial_{\rho}\phi+g^{\alpha\nu}g^{\beta\rho}\partial_{\nu}\partial_{\rho}\phi$, so that the NC part of the action in \eqref{akcija} is actually proportional to three types of terms. Namely,  terms proportional to $(\partial\phi)^2$, terms proportional to $(\partial\phi)(\partial^2\phi)$ and terms proportional to $(\partial^2\phi)^2$ (where we omit the index structure, but assume that all partial derivatives are $\partial_{\mu}=\frac{\partial}{\partial x^{\mu}}$). Therefore when calculating the equation of motion using \eqref{EL}, for the NC contribution we will get  four types of terms: $\partial\phi$, $\partial^2\phi$, $\partial^3\phi$ and $\partial^4\phi$.

It can be seen that in the special case of  time-like deformations,
$a_{\mu}=(a,\vec{0})$, for the BTZ metric and under the approximation
of the  long wavelength limit there will be no contributions in the radial equation  coming
from the higher derivative terms, as far as  the lowest order in
deformation is concerned.

The analysis presented so far is applicable for a general curved
  spacetime metric $g_{\mu\nu}(x)$. As a next step
  we use the explicit form of the BTZ metric  in eq.(\ref{akcija}) and use the $\kappa$-deformed
scalar field $\hat{\phi}$ to probe the classical BTZ  geometry in order to infer new features
  that the noncommutativity might bring into the black hole
  physics. Also the choice  $a_{\mu}=(a, \vec{0})$ is understood.
 Even with these simplifying assumptions, the equations of motion
still appear to be non-trivial. Nevertheless,  there is still enough
  space for making further simplifications, motivated and based on physical grounds.
 The first approximation that we take is the 
long wavelength limit, where we keep terms in the equations of motion that are of the lowest order in derivatives 
($\partial\phi>>\partial^2\phi,\partial^3\phi,\partial^4\phi$). In this approximation the terms dependent on $\alpha$ 
and $\gamma$  do not contribute since they are all proportional to
  $\partial^{(2,3,4)}\phi$. Consequently, 
 only terms depending on $\beta$
  give rise to noncommutative contributions and this in turn leads to
  the situation where only  terms proportional 
to  $\partial\phi$ survive.
  The outcome of the foregoing analysis is  that only realizations 
  with  nonvanishing parameter $\beta$ contribute in the lowest order of the long 
wavelength approximation. On the other side,  the choice of realization corresponds to the choice of the vacuum of the theory and this 
should be fixed by experiment, in principle. For example, $\beta=1$ corresponds to the natural realization (classical 
basis \cite{twists}).

 The final step in deriving the radial equation involves using the 
ansatz  $\phi(r,\theta,t)=R(r)e^{-i\omega t}e^{im\theta}$,  as long as $M>>1$, and keeping terms up to first order in
the deformation parameter $a$. Taking all of the above into account, we get the radial equation as
\begin{equation}
r\left(8GM-\frac{r^2}{l^2}\right)\frac{\partial^2 R}{\partial r^2}+\left(8GM-\frac{3r^2}{l^2}\right)
\frac{\partial R}{\partial r}+\left(\frac{m^2}{r}-\omega^2\frac{r}{\frac{r^2}{l^2}-8GM}-a\beta\omega\frac{8r}{l^2}\frac{\frac{3r^2}{2l^2}-8GM}{\frac{r^2}{l^2}-8GM}\right)R=0.
\end{equation}


\begin{thebibliography}{99}
\bibitem{rg}  T.~Regge and J.~A.~Wheeler,
  ``Stability of a Schwarzschild singularity,''
  Phys.\ Rev.\  {\bf 108}, 1063 (1957).

\bibitem{vish} C.V. Vishveshwara, ``Scattering of Gravitational Radiation by a Schwarzschild Black-hole'', Nature {\bf 227},936 (1970).


\bibitem{cardosoreview} Emanuele Berti, Vitor Cardoso and Andrei O. Starinets, ``Quasinormal modes of black holes and black branes'',  Class. \ Quant. \ Grav. {\bf 26}, 163001 (2009).

\bibitem{rew} 
  R.~A.~Konoplya and A.~Zhidenko,
  ``Quasinormal modes of black holes: From astrophysics to string theory,''
  Rev.\ Mod.\ Phys.\  {\bf 83}, 793 (2011)
  [arXiv:1102.4014 [gr-qc]].

\bibitem{malda} O. Aharony, S. Gubser, J. M. Maldacena, H. Ooguri and  Y. Oz, ``Large N field theories, string theory and gravity'', 
Phys.\ Rept. \ {\bf 323}, 183 (2000).
\bibitem{horo} G.~T.~Horowitz and V.~E.~Hubeny,
  ``Quasinormal modes of AdS black holes and the approach to thermal equilibrium,''
  Phys.\ Rev.\ D {\bf 62}, 024027 (2000)
  [hep-th/9909056].
\bibitem{danny1}  D.~Birmingham, I.~Sachs and S.~N.~Solodukhin,
  ``Conformal field theory interpretation of black hole quasinormal modes,''
  Phys.\ Rev.\ Lett.\  {\bf 88}, 151301 (2002)
  [hep-th/0112055].
\bibitem{danny2} D.~Birmingham, I.~Sachs and S.~N.~Solodukhin,
  ``Relaxation in conformal field theory, Hawking-Page transition, and quasinormal normal modes,''
  Phys.\ Rev.\ D {\bf 67}, 104026 (2003)
  [hep-th/0212308].
\bibitem{cones}A. Connes, Noncommutative geometry, Accademic Press, 1994.
\bibitem{dop1} S.~Doplicher, K.~Fredenhagen and J.~E.~Roberts,
  ``Space-time quantization induced by classical gravity,''
  Phys.\ Lett.\ B {\bf 331}, 39 (1994).
\bibitem{dop2} S.~Doplicher, K.~Fredenhagen and J.~E.~Roberts,
  ``The Quantum structure of space-time at the Planck scale and quantum fields,''
  Commun.\ Math.\ Phys.\  {\bf 172}, 187 (1995)
  [hep-th/0303037].
\bibitem{btzkappa} B.~P.~Dolan, K.~S.~Gupta and A.~Stern,
  ``Noncommutative BTZ black hole and discrete time,''
  Class.\ Quant.\ Grav.\  {\bf 24}, 1647 (2007)
  [hep-th/0611233].\\
 B.~P.~Dolan, K.~S.~Gupta and A.~Stern,
  ``Noncommutativity and quantum structure of spacetime,''
  J.\ Phys.\ Conf.\ Ser.\  {\bf 174}, 012023 (2009).
\bibitem{ohl} T. Ohl and A. Schenkel, ``Cosmological and Black Hole Spacetimes in 
Twisted Noncommutative Gravity'' JHEP {\bf 0910}, 052 (2009).
\bibitem{kappa} J.~Lukierski, H.~Ruegg, A.~Nowicki and V.~N.~Tolstoi,
  ``Q deformation of Poincare algebra,''
  Phys.\ Lett.\ B {\bf 264}, 331 (1991).\\
J.~Lukierski and H.~Ruegg,
  ``Quantum kappa Poincare in any dimension,''
  Phys.\ Lett.\ B {\bf 329}, 189 (1994)
  [hep-th/9310117].\\
S.~Majid and H.~Ruegg,
  ``Bicrossproduct structure of kappa Poincare group and noncommutative geometry,''
  Phys.\ Lett.\ B {\bf 334}, 348 (1994)
  [hep-th/9405107].

\bibitem{lemos} Vitor Cardoso and Jose P.S. Lemos, ``Scalar, electromagnetic and Weyl perturbations of BTZ black holes: Quasinormal modes'',  Phys. \ Rev. {\bf D 63}, 124015 (2001).

\bibitem{d1} D.~Birmingham,
  ``Choptuik scaling and quasinormal modes in the AdS / CFT correspondence,''
  Phys.\ Rev. {\bf D 64}, 064024 (2001)
  [hep-th/0101194].
\bibitem{our} 
  K.~S.~Gupta, E.~Harikumar, T.~Juric, S.~Meljanac and A.~Samsarov,
  ``Effects of Noncommutativity on the Black Hole Entropy,''
  Adv.\ High Energy Phys.\  {\bf 2014}, 139172 (2014)
  [arXiv:1312.5100 [hep-th]].
        \bibitem{sm1}  S.~Meljanac and M.~Stojic,
  ``New realizations of Lie algebra kappa-deformed Euclidean space,''
  Eur.\ Phys.\ J.\ C {\bf 47}, 531 (2006)
  [hep-th/0605133].
\bibitem{sm2}  S.~Kresic-Juric, S.~Meljanac and M.~Stojic,
  ``Covariant realizations of kappa-deformed space,''
  Eur.\ Phys.\ J.\ C {\bf 51}, 229 (2007)
  [hep-th/0702215].
\bibitem{sm3} S.~Meljanac and S.~Kresic-Juric,
  ``Generalized kappa-deformed spaces, star-products, and their realizations,''
  J.\ Phys.\ A {\bf 41}, 235203 (2008)
  [arXiv:0804.3072 [hep-th]].
\bibitem{epjc} S.~Meljanac, A.~Samsarov, M.~Stojic and K.~S.~Gupta,
  ``Kappa-Minkowski space-time and the star product realizations,''
  Eur.\ Phys.\ J.\ C {\bf 53}, 295 (2008)
  [arXiv:0705.2471 [hep-th]].
        \bibitem{twists} T.~Juric, S.~Meljanac and R.~Strajn,
  ``Twists, realizations and Hopf algebroid structure of kappa-deformed phase space,''
  Int.\ J.\ Mod.\ Phys.\ A {\bf 29}, no. 5, 1450022 (2014)
  [arXiv:1305.3088 [hep-th]].
        \bibitem{bekenstein} J.~D.~Bekenstein,
  ``Black holes and the second law,''
  Lett.\ Nuovo Cim.\  {\bf 4}, 737 (1972).
        \bibitem{Hod:1998vk}
          S.~Hod,
          ``Bohr's correspondence principle and the area spectrum of quantum black holes,''
          Phys.\ Rev.\ Lett.\  {\bf 81} (1998) 4293.
  [gr-qc/9812002].
          
\bibitem{Maggiore:2007nq}
          M.~Maggiore,
          ``The Physical interpretation of the spectrum of black hole quasinormal modes,''
          Phys.\ Rev.\ Lett.\  {\bf 100} (2008) 141301.
 [arXiv:0711.3145 [gr-qc]].
          
          
\bibitem{gross} S.~B.~Giddings, D.~J.~Gross and A.~Maharana,
  ``Gravitational effects in ultrahigh-energy string scattering,''
  Phys.\ Rev.\ D {\bf 77}, 046001 (2008)
  [arXiv:0705.1816 [hep-th]].

\bibitem{genunc} M.~Maggiore,
  ``The Algebraic structure of the generalized uncertainty principle,''
  Phys.\ Lett.\ B {\bf 319}, 83 (1993)
  [hep-th/9309034].\\
 M.~Maggiore,
  ``Quantum groups, gravity and the generalized uncertainty principle,''
  Phys.\ Rev.\ D {\bf 49}, 5182 (1994)
  [hep-th/9305163]. 

\bibitem{kowalski} J.~Kowalski-Glikman and S.~Nowak,
  ``Doubly special relativity theories as different bases of kappa Poincare algebra,''
  Phys.\ Lett.\ B {\bf 539}, 126 (2002)
  [hep-th/0203040].

\bibitem{jonke}  M.~Dimitrijevic, L.~Jonke, L.~Moller, E.~Tsouchnika, J.~Wess and M.~Wohlgenannt,
  ``Deformed field theory on kappa space-time,''
  Eur.\ Phys.\ J.\ C {\bf 31}, 129 (2003)
  [hep-th/0307149].

\bibitem{borowiecpachol} A.~Borowiec and A.~Pachol,
  ``Classical basis for kappa-Poincare algebra and doubly special relativity theories,''
  J.\ Phys.\ A {\bf 43}, 045203 (2010)
  [arXiv:0903.5251 [hep-th]].

\bibitem{sams} S.~Meljanac and A.~Samsarov,
  ``Scalar field theory on kappa-Minkowski spacetime and translation and Lorentz invariance,''
  Int.\ J.\ Mod.\ Phys.\ A {\bf 26}, 1439 (2011)
  [arXiv:1007.3943 [hep-th]].\\
 S.~Meljanac, A.~Samsarov, J.~Trampetic and M.~Wohlgenannt,
  ``Scalar field propagation in the $\phi^4$ kappa-Minkowski model,''
  JHEP {\bf 1112}, 010 (2011)
  [arXiv:1111.5553 [hep-th]].
          
\bibitem{banados} M.~Banados, C.~Teitelboim and J.~Zanelli,
  ``The Black hole in three-dimensional space-time,''
  Phys.\ Rev.\ Lett.\  {\bf 69}, 1849 (1992)
  [hep-th/9204099].\\
         M.~Banados, M.~Henneaux, C.~Teitelboim and J.~Zanelli,
  ``Geometry of the (2+1) black hole,''
  Phys.\ Rev.\ D {\bf 48}, 1506 (1993)
  [Phys.\ Rev.\ D {\bf 88}, no. 6, 069902 (2013)]
  [gr-qc/9302012].

\bibitem{kokkotas} Kostas D. Kokkotas and Bernd G. Schmidt, ``Quasi-Normal Modes of Stars and Black Holes'', Living Rev. Rel.{\bf 2}, 2 (1999), [gr-qc/9909058]. 

\bibitem{cam} http://www.ast.cam.ac.uk/research/cosmology.and.fundamental.physics/gravitational.waves.
        
\bibitem{veza}
  R.~Bufalo and A.~Tureanu,
  ``An analogy between the Schwarzschild solution in a noncommutative gauge theory and the Reissner-Nordstr\"om metric,''
  arXiv:1410.8661 [hep-th].


\bibitem{Kunstatter:2002pj}
  G.~Kunstatter,
  ``d-dimensional black hole entropy spectrum from quasinormal modes,''
  Phys.\ Rev.\ Lett.\  {\bf 90} (2003) 161301.
 [gr-qc/0212014].
  

\bibitem{vagenas} E. C. Vagenas,
  ``Area spectrum of rotating black holes via the new interpretation of quasinormal modes,''
  JHEP {\bf 0811}, 073 (2008)
  [arXiv:0804.3264 [gr-qc]].
        
\bibitem{medved} A.~J.~M.~Medved,
  ``On the Kerr Quantum Area Spectrum,''
  Class.\ Quant.\ Grav.\  {\bf 25}, 205014 (2008)
  [arXiv:0804.4346 [gr-qc]].
        
\bibitem{sullivan} D. Sullivan, in Proceedings of the 1978 Stony Brook Conference on Riemann Surfaces and
Related Topics, edited by I. Kra and B. Maskit, Annals of Mathematics Studies No. 97,
Princeton University Press, Princeton, New Jersey, 1981; C. McMullen, Bull. Am. Math.
Soc. 27, (1992) 207; Invent. Math. 99, (1990) 425.

\bibitem{sendanny} D. Birmingham, C. Kennedy, Siddhartha Sen and Andy Wilkins, ``Geometrical finiteness, holography, and the BTZ black hole'',  Phys. \ Rev. \  Lett. {\bf 82}, 4164 (1999).

\bibitem{senksg1} Kumar S. Gupta and Siddhartha Sen, ``Geometric finiteness and non-quasinormal modes of the BTZ black hole'' Phys. \ Lett. {\bf B 618}, 237 (2005). 

\bibitem{senksg2} Kumar S. Gupta, E. Harikumar, Siddhartha Sen and M. Sivakumar, ``Geometric Finiteness, Holography and Quasinormal Modes for the Warped AdS3 Black Hole'' Class. \ Quant. \ Grav. {\bf 27}, 165012 (2010).

\bibitem{dil} I.~Sakalli,
  ``Quantization of rotating linear dilaton black holes,''
  Eur.\ Phys.\ J.\ C {\bf 75}, no. 4, 144 (2015)
  [arXiv:1406.5130 [gr-qc]]. 

\bibitem{Wei:2009yj} 
  S.~W.~Wei, R.~Li, Y.~X.~Liu and J.~R.~Ren,
  ``Quantization of Black Hole Entropy from Quasinormal Modes,''
  JHEP {\bf 0903}, 076 (2009).
[arXiv:0901.0587 [hep-th]].
        
        


        
        
\end{thebibliography}
\end{document}